\documentclass[a4paper,12pt]{spieman}  
\usepackage{amsmath,amsfonts,amssymb,bm} 
\usepackage{soul}
\usepackage{graphicx} 
\usepackage{tocloft}
\usepackage[outdir=./]{epstopdf}
\usepackage{subcaption}
\usepackage{wrapfig}

\title{NICSPol: A Near Infrared polarimeter for the 1.2 m telescope at Mount Abu Infrared Observatory}

\author[a,b,*]{Esakkiappan Aarthy}
\author[a,b]{Archita Rai}
\author[a]{Shashikiran Ganesh}
\author[a]{Santosh V. Vadawale}
\affil[a]{Physical Research Laboratory, Ahmedabad, Gujarat, India, 380009}
\affil[b]{Indian Institute of Technology, Gandhinagar, Gujarat, India, 382355}

\cftpagenumbersoff{figure}  
\cftpagenumbersoff{table} 

\begin{document} 
\maketitle

\begin{abstract}
NICSPol is a near infrared imaging polarimeter developed for the Near Infrared Camera and Spectrograph 
(NICS), one of the back end instruments of the 1.2 m Cassegrain telescope at the Mount Abu Infrared 
Observatory (MIRO), India. The polarimeter consists of a rotating wire grid polarizer which is mounted 
between the telescope optics and NICS. The polarimetric observations are carried out by rotating the 
polarizer using a motorized mechanism to determine the Stokes parameters, which are then converted into 
the polarization fraction and polarization angle. Here we report the details of the instrument and the 
results of observations of IR polarimetric standards. A set of polarized and unpolarized standards 
were observed using NICSPol over J, H and Ks bands covering 0.8 to 2.5 $\mu$m.
The observations of polarized standards using NICSPol show that, NICSPol can constrain 
polarization within $\sim$1\% for sources brighter than $\sim$16 magnitude in JHKs bands.
NICSPol is a general purpose instrument which could be used to study variety of astrophysical sources 
such as AGNs, Pulsars, XRBs, Supernovae, star forming regions etc.
With few NIR polarimeters available world-wide so far, NICSPol would be the first imaging NIR polarimeter in India. 
\end{abstract}


{\noindent \footnotesize\textbf{*}Esakkiappan Aarthy,  \linkable{aarthy@prl.res.in} , \linkable{aarthy.e@iitgn.ac.in} }


\section{Introduction}
\label{sect:intro}  
Observational astronomy revolves around collection of electromagnetic radiation from various 
celestial sources. Spectroscopy, imaging and timing are the three most popular and 
commonly used techniques to analyze and interpret the photons received. There is a fourth 
comparatively less explored but important tool: astronomical polarimetry. 
Polarization is an inherent property of electromagnetic radiation 
which indicates the dominant orientation of the electric (\& magnetic) field vector.
Polarimetry gives two additional parameters: polarization angle and 
degree of polarization. 
Degree of polarization is the fraction at which the source is polarized. 
The orientation of the plane in which the electric vector oscillates, 
as it propagates, is the polarization angle.
In general, astronomical sources are unpolarized unless there exist some 
asymmetry and/or anisotropy either in terms of matter 
or field, in the source or along the path of propagation of the photons.
Hence measuring polarization helps in studying the 
magnetic field and geometry of various astronomical sources and their surrounding medium.
Despite the benefits, polarimetry is relatively less popular compared to spectroscopy 
and imaging due to certain inherent issues.
The prime reason is that most of the sources possess very low polarization (a few \%) 
and hence longer exposure times are needed compared to spectroscopy. \\ 

The origin of polarization mostly differ over different parts of the 
electromagnetic spectrum.
Polarization in UV, optical and IR are most commonly due to 
scattering by dust grains and magnetic field 
that orients the dust grains. Non thermal radiation from 
different sources are also polarized. 
In X-ray regime depending on the photon energy, geometry and magnetic field, polarization 
could be due to scattering (Rayleigh, inverse Compton), cyclotron, synchrotron \cite {aitken88} 
or due to more exotic physical processes like vacuum polarization and birefringence through extreme
magnetic fields.\cite {gnedin78}\textsuperscript{,}\cite {meszaros80} 
In radio regime also non thermal 
radiation is synchrotron.
The principle and methodology to acquire and analyze polarized data vary 
over the different regions of the spectrum. 
Assuming the initial radiation from the internal structure of the source to be unpolarized, 
measuring polarization is the only way to know about the point of interactions 
which polarizes it.
The study of polarization induced by scattering is a powerful technique to understand 
the intervening matter and field distribution, when the original radiation 
is expected to be unpolarized (e.g. thermal radiation).\\

The Infrared regime of the EM spectrum helps in unveiling many embedded systems 
which could not be seen otherwise at optical wavelengths.
The first measurement of IR polarization dates back to 1975 \cite {martin75} and since then 
there have been several ground and space based instruments covering different 
regions from $\sim$0.8 $\mu$m in near IR to few tens of $\mu$m in the far IR 
studying wide range of astronomical sources.
Few well known near Infrared polarimeters are: SIRPOL at the 1.4 m IRSF \cite {kandori06}, Mimir at the 1.8 m Perkins 
telescope \cite {clemens07}, NICMOS at the 2.4 m HST \cite {schultz03}, MMTPOL at the 6.5 m MMT Observatory 
\cite {jones07}, SOFI at the 3.58 m NTT telescope, 
La Silla Observatory \cite {wolf02}, POLICAN at the 2.1 m telescope of the Guillermo Haro Astrophysical Observatory 
(OAGH) located in Cananea, Sonora, Mexico \cite {devaraj18}, TRISPEC as a visitor instrument at several facilities 
including UKIRT, UH, OAO, Subaru telescope \cite {watanbe05}, SPHERE at the 8.2 m VLT \cite {beuzit13}, 
HONIR at the 1.5-m Kanata telescope \cite {akitaya14} etc.\\

We have recently added a polarimeter NICSPol to the existing 
Near Infrared Camera and Spectrograph (NICS) which is one of the back end instruments 
for the 1.2 m Cassegrain f/13 telescope 
located at the Mount Abut Infrared Observatory of Physical Research Laboratory 
\cite{anandrao08}.
This makes NICSPol the first imaging IR polarimeter in India.
NICS is capable of doing both photometry and spectroscopy covering a wide range of 0.8 to 2.5 $\mu$m
(Y, J, H, Ks) and has been serving as a work horse for the past several years to study
diverse objects like AGN, galaxies, supernovae, novae and compact objects etc.
Kolokolova et al, 2015 \cite {kolokolova15} and other excellent review articles \cite{trippe14}, 
\cite{snik13}, cover polarimetry and the scientific aspects over a wide electromagnetic domain.
Polarization of light in infrared could be majorly due to 
scattering (e.g. polarization in comets, planets, asteroids etc are due to scattering 
by dust grains), and/or non thermal radiation (e.g. synchrotron radiation from AGN, XRBs).
The scientific prospects of IR polarization in case of both galactic and extra galactic sources 
have been extensively reported in the literature \cite {hines09}, \cite{hoffman09}, \cite{clemens09}, 
\cite{clemens09gan}, \cite {clemens12}.
Having optical polarization measurement with the 50 cm telescope of PRL \cite {ganeshs13}, and the
hard X-ray polarization capability of AstroSat CZTI \cite {vadawale15}, NICSPol would
provide a fantastic opportunity for simultaneous polarimetry over a wide range of EM spectrum.\\

Here we report the NICSPol instrument design and its calibration results for 
IR polarimetric standards. 
Different polarization measurement techniques are briefed in Section \ref {sec:2} and 
NICSPol instrument specifications are given in Section \ref {sec:3}.
Sections \ref {sec:4} and \ref {sec:5} describe the observations 
and the data reduction. The results are mentioned in Section \ref {sec:6} followed by 
the summary in Section \ref {sec:7}.

\section{NIR Polarimetry}
\label{sec:2}
\subsection{Instrumental Techniques}
The means of acquiring the polarimetric information varies depending on the energy regime of the incoming 
photons.
There are several methods to measure polarized radiation from astronomical sources. 
The efficiency of the measurement method is one of the deciding factors in preferring a 
particular scheme. Using a Wollaston prism along with a rotating half wave plate modulator, gives both the 
orthogonal 
polarization states in a single exposure, hence also building up on the total efficiency of the system 
\cite {wpp}. 
WeDoWo (Wedged Double Wollaston) is another similar technique with the difference that two Wollastons 
are combined together, giving images at four angles (0\textsuperscript{o}, 45\textsuperscript{o}, 90\textsuperscript{o}, 
135\textsuperscript{o}) simultaneously in a single exposure \cite {wdw}. 
This technique, hence, helps in simultaneous observation and prevents variations in the sky background during the 
exposure, from affecting the results. The third one and the one used in NICSPol is a Wire-grid 
Polarizer (WGP).
It works on the principle of dichroism, and functions as an absorptive polarizer. 
The selective direction in the case of WGP is perpendicular to the wire-grids, resulting 
in linear polarization. The wavelength range being covered by a WGP is 
dependent on the spacing and widths of the wires in the WGP \cite {wgp}. The extinction ratio of polarizers is a measure 
of the ability to attenuate light in direction perpendicular to transmission axis of 
polarizer\cite {thorlabs}.This is an 
important parameter which decides the efficiency of the polarizers. This ratio for WGP in infrared has 
a value of 1000:1. Compared to other polarization optics components, WGP can be made in larger apertures. 
Thus, WGPs are effective polarizers with large field of view (FOV), are fairly compact and have 
good stability \cite {wgp}. Hence, they are used extensively in polarization measurement.
Our choice for WGP method of polarization measurement in NICSPol is also motivated by the limited space 
available in the light path between the telescope Cassegrain mounting flange and the instrument entrance aperture.

\subsection{Analysis Technique}

The polarization of a source is defined by two parameters, the polarization fraction (PF) 
or the degree of polarization and the polarization angle (PA). 
There are different methods to acquire these two information among which 
the most preferentially used method in optical or IR is the Stokes method.
When a polarized beam passes through an analyzer which is rotating at discrete steps, 
the output beam gets modulated and follows a Cos2$\theta$ distribution.  
A mathematical fit (equation \ref {eqn:fit}) to the observed intensities provides the PF and PA.
But getting intensities of a source at multiple steps over 0\textsuperscript{o} to 180\textsuperscript{o} 
is quite time consuming and in the case of the infrared domain the assumption that the sky has not 
changed may not hold.
The advantage of Stokes method is that, it requires intensities  
only at 3 (or 4) different orientations and effectively provides the PA and PF using any 3 (or all 4) of the 
I\textsubscript{0}, I\textsubscript{45}, I\textsubscript{90}, I\textsubscript{135} 
(intensities obtained by rotating the WGP in steps of 45\textsuperscript{o}) measurements.
Given by GG Stokes in 1852 \cite{stokes52} and introduced in astronomy by 
S Chandrasekhar in 1947, Stokes parameters I, Q, U, V describe the polarization state 
of a system in terms of intensities \cite{chandra60}. I (equation \ref {eqn:I}) represents the total intensity, 
Q (equation \ref {eqn:Q})  and U (equation \ref {eqn:U})
represent linear polarization and V represents circular polarization.
The Stokes parameters are calculated from the observations of flux measured at different 
orientations of the WGP.
The following equations give the relation between intensity and Stokes parameters \cite {frecker76}.

\begin{equation}
\label {eqn:I}
{I	=	\frac{1}{2} (I_{0}+I_{45}+I_{90}+I_{135}) }\ \  
\end{equation}
\begin{equation}
\label {eqn:Q}
{Q	=	{I_{0}-I_{90}}} \ \  
\end{equation}
\begin{equation}
\label {eqn:U}
{U	=	{I_{45}-I_{135}}} \ \  
\end{equation}
Alternatively, Stokes parameters could be obtained by by fitting equation \ref {eqn:fit},  
\begin{equation}
\label {eqn:fit}
{I_{j} =	\frac{1}{2}[I_{0} \pm Q cos2 \theta_{j} \pm U sin2 \theta_{j} ]} \ \  
\end{equation}
The PF and PA are obtained from the Stokes parameters using equations \ref {eqn:PF} and \ref {eqn:PA}.
\begin{equation}
\label {eqn:PF}
{PF	=	\sqrt{\frac{Q^{2}+U^{2}}{I}}} \ \  
\end{equation}
\begin{equation}
\label {eqn:PA}
{PA	=	\frac{1}{2} tan^{-1}\frac{U}{Q}} \ \  
\end{equation}


\section{NICSPol Instrument description}
\label{sec:3}
\begin{figure}[!ht]
\begin{center}
\begin{tabular}{c}
    \includegraphics[width=0.9\linewidth]{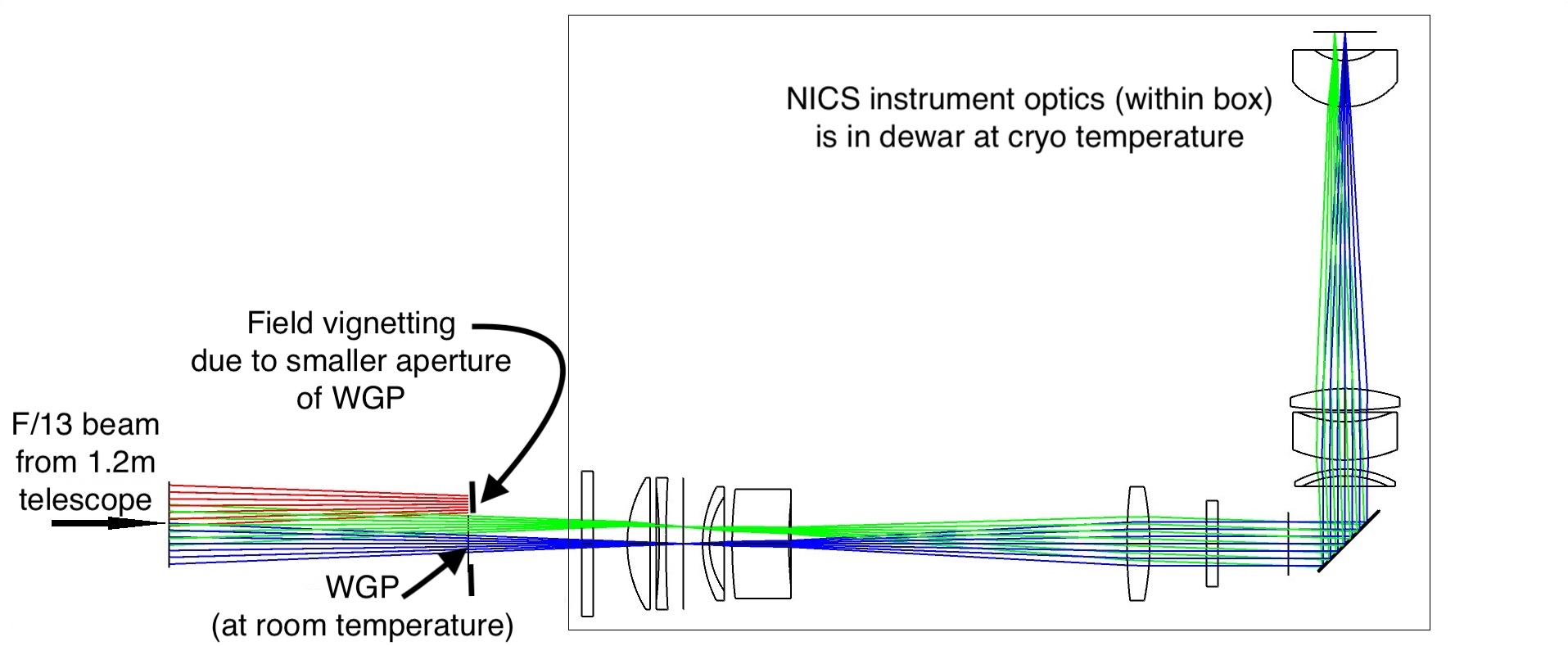}
\end{tabular}
\end{center}
    \caption
{\label{fig:1} NICSPOL instrument with the WGP placed before the Cryostat window of NICS. 
From the ray diagram, its evident that the WGP position and size obscures part of the rays 
to reach the detector. The full FOV as covered by NICS, is thus reduced in NICSPOL 
(vignetting as seen by the extreme field rays not passing through the WGP to reach the detector). 
The different field projections on the detector plane are shown by rays of different colors.} 
\end{figure}

\begin{figure}[!ht]
\begin{center}
\begin{tabular}{c}
    \includegraphics[width=0.8\linewidth]{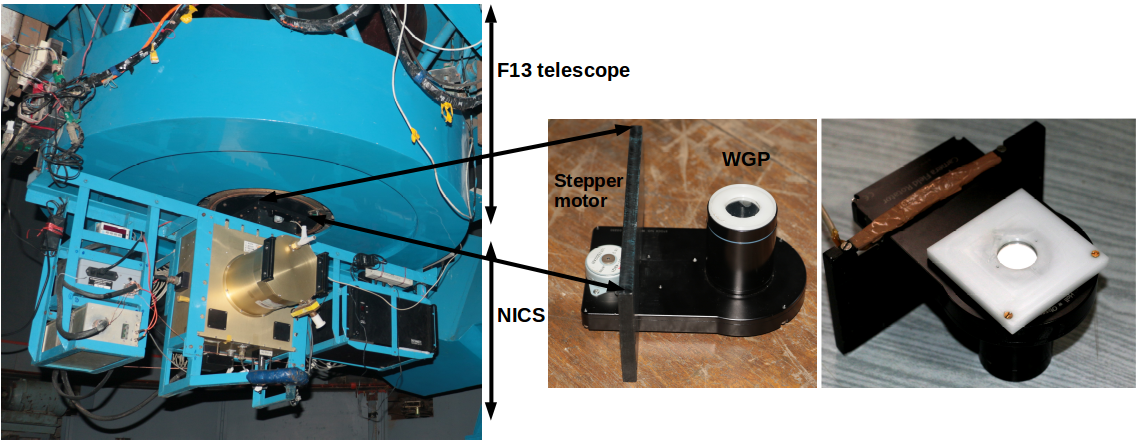}
\end{tabular}
\end{center}
    \caption
{\label{fig:2} (a) NICS mounted on the f/13 telescope at MIRO along with the polarizer unit. 
(b) NICSPol unit with the 25 x 25 cm WGP (c) The second WGP mounted in NICSPol 
to polarize the incoming light to achieve 100\% polarized radiation.} 
\end{figure}

NICSPol consists of a 25.0 x 25.0 mm WGP 
(WGP) commercially available from Thorlabs, 
which covers a wavelength range of 250 nm to 4 $\mu$m.
The layout of the NICSPol instrument is shown in Figure \ref {fig:1}. Figure \ref {fig:2} shows the instrument mounted on 
the Cassegrain plate of the MIRO telescope. The WGP mounting details are also shown in this figure.
NICS is capable of doing both photometry and spectroscopy by the use of mirror and grating.
The specifications of NICS are given in Table \ref {tab:1}.
NICS consists of a filter wheel with Y, J, H, K, Ks filters. 
The detector in NICS is Teledyne H1RG detector with 1024 x 1024 arrays which is cooled 
by using liquid Nitrogen and covers a FOV of 8 x 8 sq arc min \cite{anandrao08}.
At the top of NICS, the part which is attached to the instrument ring of the telescope is 
a 20 cm x 20 cm square box which contains a beam splitter. This essentially sends a 
fraction of the light from the source to a guiding ccd which helps in telescope tracking. 
The use of a warm WGP is necessitated by the requirement that the polarization module could be inserted in (or removed from) 
the light path without fully dismounting the entire instrument.  The only drawback in this scheme (warm WGP) is the 
increased thermal background in the K band which restricts us to use only the Ks filter in the 
longer wavelength side.\\
 
\begin{figure}[!ht]
\begin{center}
\begin{tabular}{c}
    \includegraphics[width=0.5\linewidth]{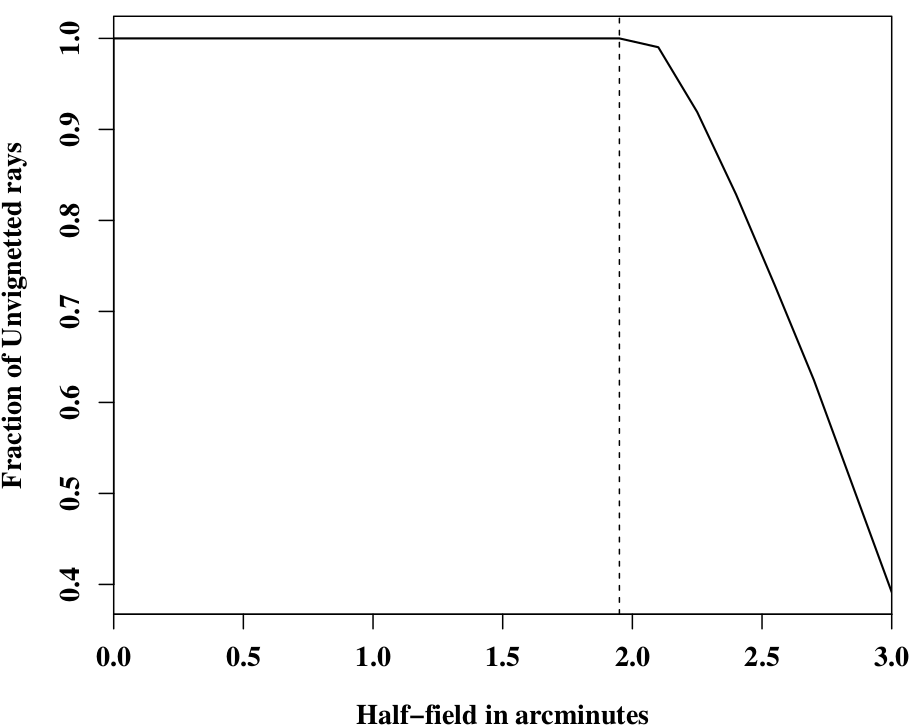}
\end{tabular}
\end{center}
    \caption
{\label{fig:3} Field vignetting caused by the introduction of the WGP.   It is seen that the vignetting 
starts beyond the dotted line (at 1.95 arcmin).} 
\end{figure}

An Optec Pyxis LE Camera Field Rotator was found suitable to mount the polarizer. The field rotator has a 
barrel which is rotated by a stepper motor with a step size of 1\textsuperscript{o} rotation. 
A delrin module was made to hold the polarizer firmly and the module was 
mounted in the barrel of the field rotator.
A side of the beam splitter holder box was replaced by the module 
as in Figure \ref {fig:2} at a proper position so that the polarizer is 
exactly along the line through which the light from the secondary reaches the NICS optics. 
The polarimetry and imaging/spectroscopic modes are easily interchangeable.
Since the size of the WGP is such that it blocks some of the incoming beam, there is strong vignetting 
(see Figure \ref {fig:3}) beyond half-field of 1.95 arcmin.
This limits the useful FOV to 3.9 arc min (dia).\\

The faintest stars seen using NICS in J, H, Ks bands, using our 1.2 m telescope reach magnitudes of $\sim$17 with 
individual exposures of 30 to 40 sec per frame and a net exposure of 30 minutes. However, the introduction 
of the WGP will result in brighter detection limits, due to the $\sim$85\% transmission of the WGP in 
NIR \cite {thorlabs} making the NICSPol limiting magnitude $\sim$16.8. 
However in polarimetric mode we expect that these magnitudes might be achieved only under very stable skies. 

\section{Observations}
\label{sec:4}

This section describes the list of observations made to verify the instrument performance 
using lab and sky polarimetric standards.
The instrument had to be checked with polarized and unpolarized standards for efficiency and 
instrumental polarization respectively.
NICSPol was calibrated using both polarized and unpolarized stellar standards.
A second WGP of same dimensions was also used to achieve 100 \% polarized light. 
This WGP was mounted stationary in NICSPol in a way that light from the source would first pass through it, 
get polarized and pass through the rotating WGP resulting in a 100 \% polarized light. 
For calibration with stellar sources both isolated stars and crowded fields were observed. \\

Table \ref {tab:2} shows the complete log of all 
the observations made for the calibration of NICSPol.
A major issue faced was the small number of stellar polarimetric standards in NIR. 
The few available standards are polarized to a maximum 
of 3 - 4\% in J band.  
6 UKIRT polarimetric standards : HD 283809, HD 204827, HD 29333, HD 283725, HD 283855 and HD 283701
and four unpolarized standards : HD 202573, HD 212311, HD 103095 and HD 65583 (to check for instrumental polarization)
\cite {whittet92}\textsuperscript{,}\cite {serkowski74}
were selected based on their availability 
at MIRO winter sky. The polarized standards were chosen such that they 
possess a degree of polarization of about 2 - 4 \% and these were followed up over multiple nights from 
September to November 2017 and December 2018 to January 2019.
A critical part of NIR observations is the removal of sky contribution, which is bright in infrared. 
For a particular source, for each filter 
the polarimetric images were taken by dithering the source in 
minimum 5 positions and the sky frame was made from these multiple dithered frames of the source itself. 
In case of extended sources sky frames were taken separately by taking 
the source out of the field of detector for the same exposure as the source, 
assuming the sky to be constant over this dithering period. 
The source images were taken for 3 filters x 5 dithering positions (with 3 frames per position) x 4 orientations of the WGP.
The exposure time varied from 801 ms 
in case of bright point sources (minimum achievable by NICS) to 20 s each for faint sources.
Also to check how closely the results from 4 orientations would match with the result from data with multiple points 
fitted using equation \ref {eqn:fit}, 
HD 56591 was observed with 2 WGPs (i.e. polarizing the light from the source), 
for 0 \textsuperscript{o} to 150 \textsuperscript{o} with 15 \textsuperscript{o} angular step size.

\section{Data Reduction and Analysis}
\label{sec:5}

As explained in Section \ref {sec:4} the Stokes method is followed to analyze the polarimetric data. 
The analysis was carried out using standard \textit{IRAF} (Image Reduction and Analysis Facility)
procedures along with few \textit{IDL} scripts to ease the analysis.
The initial steps to analyze the polarized images in IR are the same as standard IR 
photometry.
For a given source and a given band, the images had been taken by rotating the polarizer in
four orientations, 0\textsuperscript{o}, 45\textsuperscript{o}, 90\textsuperscript{o}, 135\textsuperscript{o} 
and multiple frames are taken for each dithering position. 
And as mentioned earlier the IR sky is bright hence sky subtraction is a critical step in 
the analysis. 
Sky frames were constructed by median combining source frames taken at different dithering positions.
This sky image obtained was subtracted from all the raw source images to get the sky subtracted 
frames. To improve the S/N ratio multiple frames of all the positions should be added up. 
Since the position of the source would be different in each frame, it was shifted to a common point 
with respect to a particular star and the shifted images were 
combined to give a final image for each orientation. 
The next step is to get the intensities of the source at the 4 orientations. 
Similar to photometry the magnitude of the source should be extracted and then converted 
into intensity. 
For this the \textit{IRAF phot} procedure was used to do aperture photometry, 
where a range of apertures were defined for the 
source and using the curve of growth plot a suitable aperture size was considered for 
further analysis. 
The above procedures were repeated for all 4 orientations of observations hence giving 4 intensities 
I\textsubscript{0}, I\textsubscript{45}, I\textsubscript{90}, I\textsubscript{135}.
The Stokes parameters were calculated from these intensities and the degree of polarization 
and the polarization angle are obtained from the Stokes parameters.
The additional advantage of following Stokes method is that since normalization is done 
with Intensity (I) any systematic error in the parameters are cancelled out. 
In the case of extended sources a standard \textit{IRAF}
procedure, linpol, would directly give pixel by pixel information of PF, PA, I, Q, U.

\section{Results}
\label{sec:6}

\subsection{100 \% Polarized Light}
\label{sec:6.1}
\begin{figure}[!ht]
\begin{center}
\begin{tabular}{c}
    \includegraphics[width=0.9\linewidth]{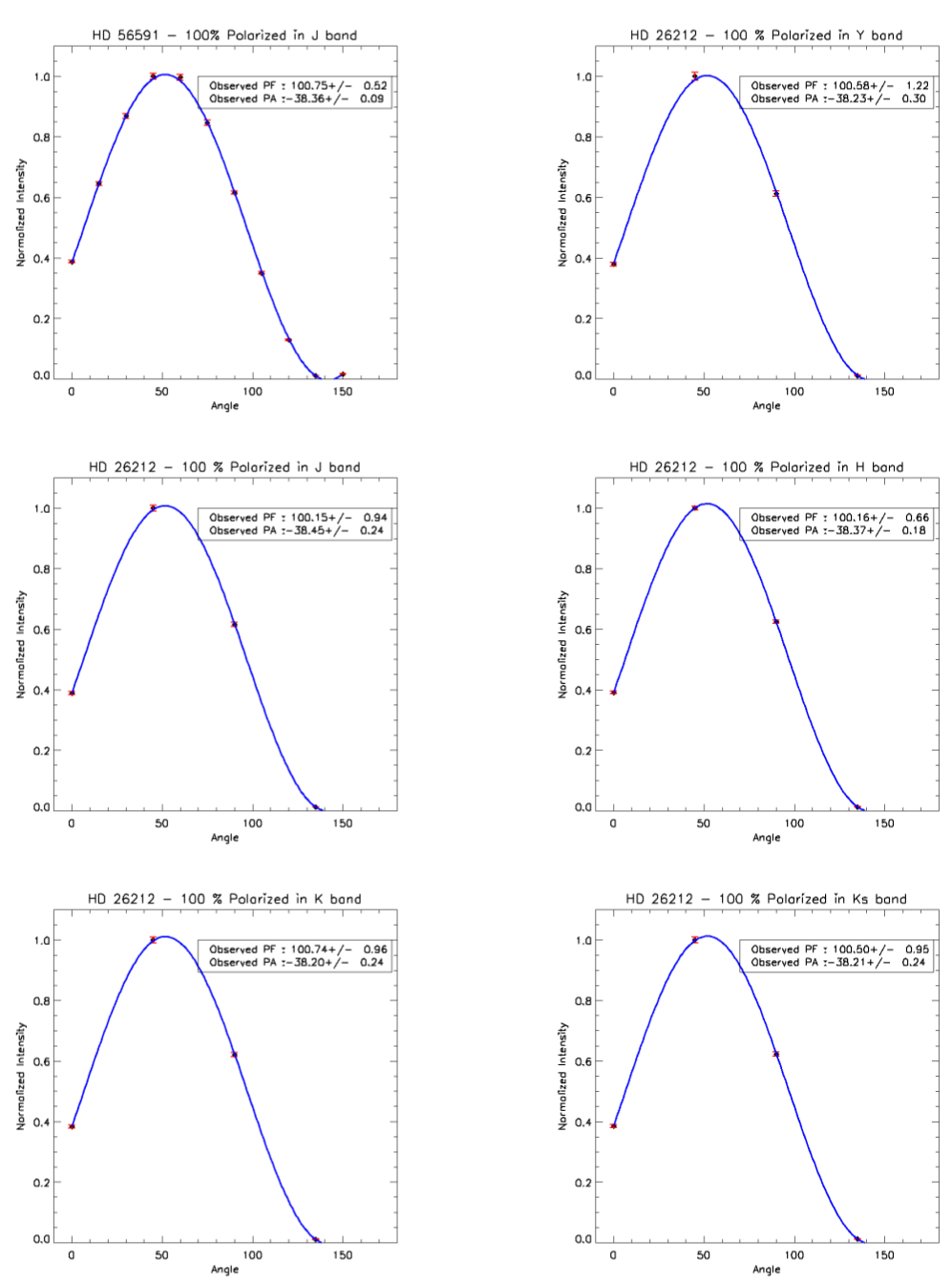}
\end{tabular}
\end{center}
    \caption
{\label{fig:4}  Modulation curves for 100 \% polarized light for sources (a) HD 56591 in J band 
and (b)-(f) HD 26212 in Y, J, H, K, Ks bands with fitted 
PF and PA quoted inside.} 
\end{figure}

\begin{figure}[!ht]
\begin{center}
\begin{tabular}{c}
    \includegraphics[width=0.5\linewidth]{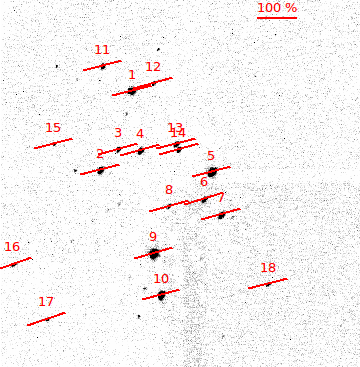}
\end{tabular}
\end{center}
    \caption
{\label{fig:5} 100\% polarized light using second WGP from NGC 2548, marked with 
polarization vectors using the PF and PA obtained. }
\end{figure}

By mounting a second stationary WGP the incoming light from a source is polarized, 
using this the polarimetric efficiency of NICSPol was tested by observing 
HD 56591, HD 26212, NGC 2548 and M3.
The obtained PF and PA are tabulated in Tables \ref {tab:3} and \ref {tab:4}.
The modulation curves for HD 56591 and HD 26212 are plotted in Figure \ref {fig:4}. 
The reason behind the PF crossing 100\% within the error bar is that, 
the intensity of the source at 135\textsuperscript{o} and above is very less which increases the uncertainty in 
the photometric measurement. The fitted PA of both the sources are found to be consistent over all the 
bands and the error in PA is found to be only $\sim$ 0.1 - 0.2\textsuperscript{o}. 
Figure \ref {fig:5} shows the NGC 2548 field marked with polarization vectors of all the sources 
whose details are tabulated in Table \ref {tab:4}. 
It could be seen that the results are consistent over the entire FOV of the NICSPol module over the detector.

\subsection{Unpolarized Stars}
\label{sec:6.2}
\begin{figure}[!ht]
\begin{center}
\begin{tabular}{c}
    \includegraphics[width=0.5\linewidth]{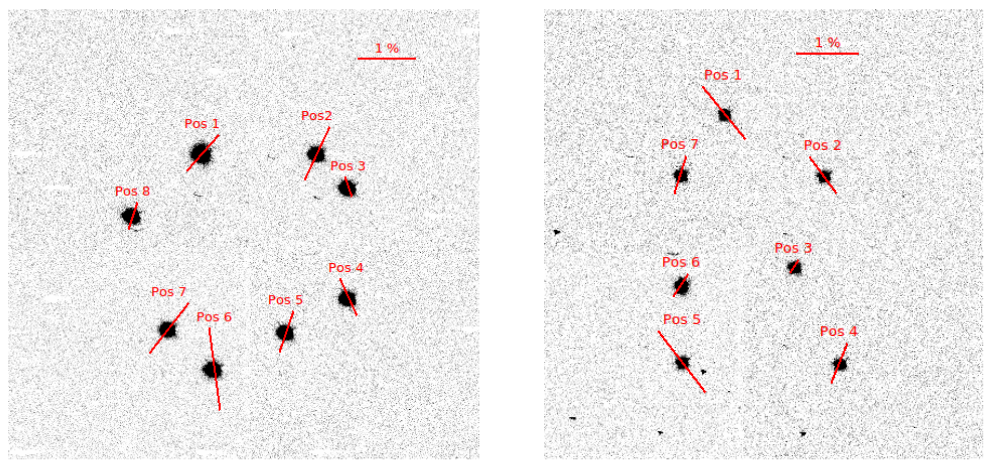}
\end{tabular}
\end{center}
    \caption
{\label{fig:6} Unpolarized standards (a) HD 103095 and (b) HD 65583 over different positions in the FOV 
marked with polarization vectors using the PF and PA obtained. }
\end{figure}

\begin{figure}[!ht]
\begin{center}
\begin{tabular}{c}
    \includegraphics[width=0.5\linewidth]{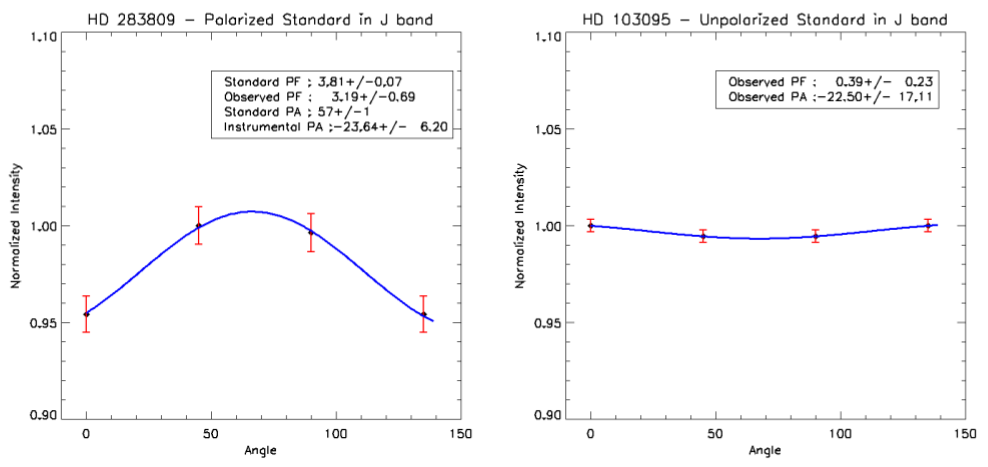}
\end{tabular}
\end{center}
    \caption
{\label{fig:7} Modulation curves fitted over 4 phase angles for (a) polarized standard HD 283809 and 
(b) unpolarized standard HD 103095.} 
\end{figure}

\begin{figure}[!ht]
\begin{center}
\begin{tabular}{c}
    \includegraphics[width=0.5\linewidth]{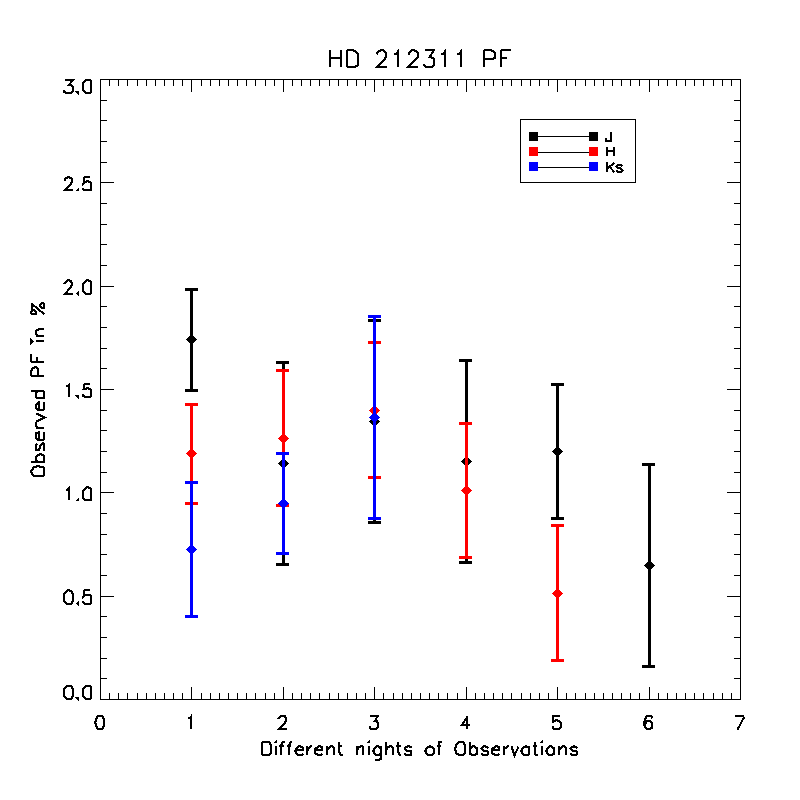}
\end{tabular}
\end{center}
    \caption
{\label{fig:8}  Observed PF for an unpolarized source HD 212311 over J, H, Ks bands 
  at different nights.} 
\end{figure}

Four unpolarized standards - HD 202573, HD 212311, HD 103095 and HD 65583 were observed using NICSPol 
and the obtained PF are tabulated in Table \ref {tab:5}. 
HD 103095 and HD 65583 was also observed by dithering the source at multiple 
positions (8 positions).
Figure \ref {fig:6} shows the PF and PA obtained over various positions in the FOV for both HD 103095 and HD 65583.
The PFs at all the positions were lesser than 1\%. and also the PAs over these positions were found to be random.
Figure \ref {fig:7} shows the modulation of unpolarized standard HD 103095.
Figure \ref {fig:8} shows the PF of one unpolarized standard HD 212311 observed over several nights which is found to be around 
1\%. From unpolarized standards it is seen that the uncertainty is around $\sim$1\%. 

\subsection{Polarized Standard Stars}
\label{sec:6.3}
\begin{figure}[!ht]
\begin{center}
\begin{tabular}{c}
    \includegraphics[width=0.5\linewidth]{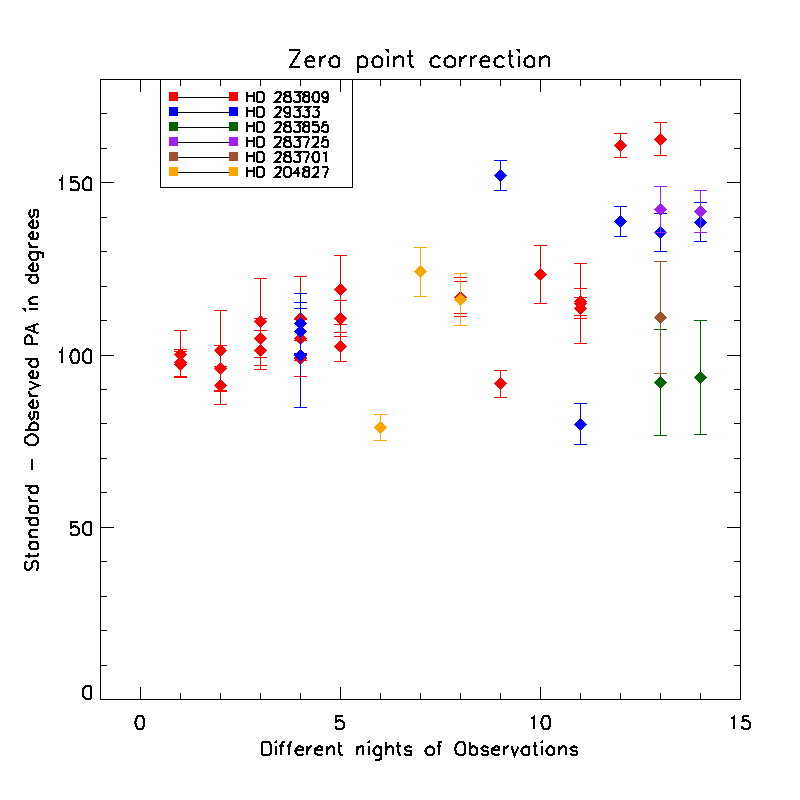}
\end{tabular}
\end{center}
    \caption
{\label{fig:9} Difference between Standard PA and instrumental PA over different observation nights. }
\end{figure}

\begin{figure}[!ht]
\begin{center}
\begin{tabular}{c}
   \includegraphics[width=0.5\linewidth]{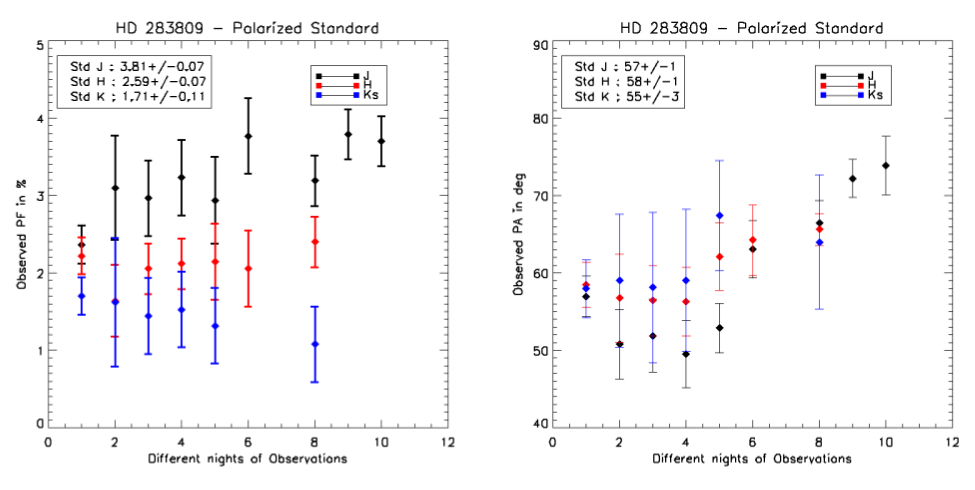}
\end{tabular}
\end{center}
    \caption
{\label{fig:10} Observed (a) PF and (b) PA for a polarized source HD 283809 over J, H, Ks bands 
at different nights with standard values quoted inside. }
\end{figure}

\begin{figure}[!ht]
\begin{center}
\begin{tabular}{c}
    \includegraphics[width=0.5\linewidth]{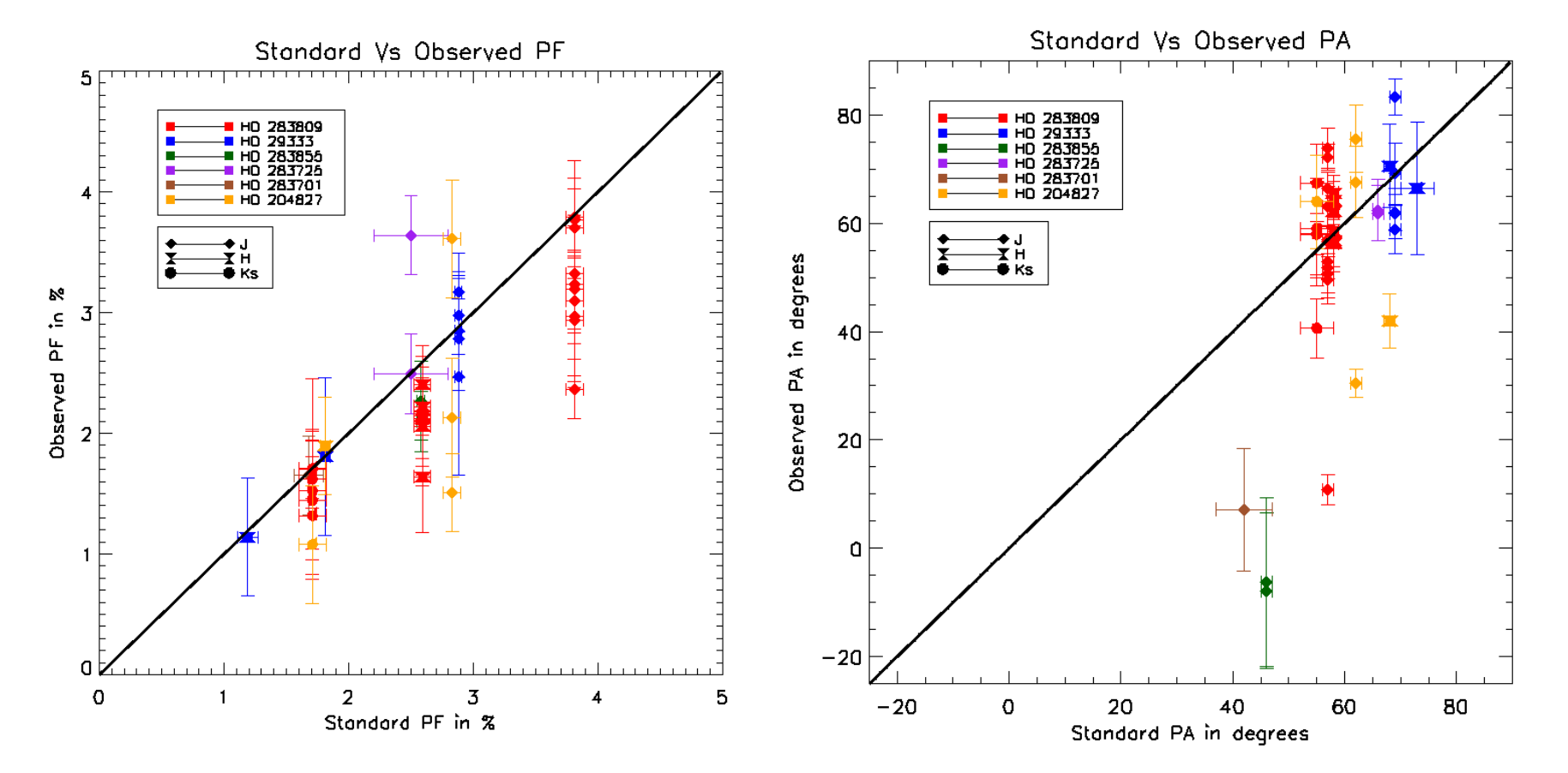}
\end{tabular}
\end{center}
    \caption
{\label{fig:11} (a) Observed PF Vs Standard PF and (b) Observed PA Vs Standard PA values for all polarized standards 
(color coded) at different nights. Different symbols are used for J, H and Ks bands as labelled inside. }
\end{figure}

IR polarized standards are less in number and the degree of polarization is generally quite low 
compared to optical wavelengths.
Six polarized standards - HD 283809, HD 204827, HD 29333, HD 283725, HD 283855 and HD 283701, 
were observed and few of these were followed up during 
several observing nights. 
The obtained PF and PA along with 
standard reported values are tabulated in Table \ref {tab:6}.
The references of reported PF, PA of chosen standards are given in Ref column.
The instrumental polarization angle obtained by fitting Equation \ref {eqn:fit} 
has to be converted into the standard polarization 
angle with reference to the local North in the equatorial coordinate system.
Over a given observation night the difference between the instrumental PA and Standard PA is expected to be same for different sources and 
by implementing this correction factor Observed PA is obtained.
Figure \ref {fig:9} shows the difference between the instrument and standard PA for all the sources 
over all observation nights. 
It could be seen that the observed values match with the 
standard values well within the error bar for sources HD 283809, HD 204827, HD 29333 and HD 283725.
Figure \ref {fig:10} shows the PF and PA values obtained for HD 283809 
with the standard values quoted inside for J, H and Ks bands and modulation curve of one observation is given in 
Figure \ref {fig:7}. 
The observed PF and PA for HD 283809 and HD 29333 were found to be consistent when focused over 
different positions of the detector FOV.
For sources HD 283855 and HD 283701 although the Observed PF match the Standard PF, the Observed PAs 
are different from the reported values. 
It should be noted that at least for HD 283701, different PA values of
42\textsuperscript{o}$\pm$5\textsuperscript{o} (Whittet et al. 1992 \cite {whittet92}) to 
33\textsuperscript{o}$\pm$1.3\textsuperscript{o} (Whittet et al. 2001 \cite {whittet01}) are
reported in literature. This indicates the possibility that the PA change could be real, 
which can only be confirmed by further observations that we plan to continue. 
Figure \ref {fig:11} shows the obtained PF and PA plotted with respect to the reported values of PF PA. 
MIMIR instrument reported the PF and PA of standard stars with small 
error bars \cite {clemens07}. 
The reason for our relatively large errors in comparison with MIMIR is, in MIMIR's case the 
observations are taken for 32 half wave plate position angles, 
i.e. for every 11.25\textsuperscript{o}. 
Also MIMIR is at 1.8 m Perkins telescope hence has large number of photons available therefore 
less photon noise. 
These result in higher precision in the measurements.   
For e.g. in one of our 100\% polarization observations (\ref {sec:6.1}), one set of observations 
were made for every 
15\textsuperscript{o} position angle covering 0\textsuperscript{o} - 165\textsuperscript{o} 
(Figure \ref {fig:4}). From these 11 position angles of the WGP PF = 100.75\%$\pm$0.52\%, 
PA = -38.36\textsuperscript{o}$\pm$ 0.09\textsuperscript{o} are obtained. 
Whereas when only 4 position angles from 
these with a stepsize of 45\textsuperscript{o} was used the values obtained are PF = 100.8\%$\pm$0.9\%, 
PA = -38.49\textsuperscript{o}$\pm$0.2\textsuperscript{o}. The increase in errors due to the 
smaller number of position angles is evident in the case of 100\% polarized light and 
this would affect the sources with low polarization. Hence in the case of NICSPol it is 
preferable to go for more position angles while observing sources with low PF, while 4 positions 
angles would be sufficient to study highly polarized sources.

\subsection{Polarization Measurement of a Photometric Standard Field}
\label{sec:6.4}
\begin{figure}[!ht]
\begin{center}
\begin{tabular}{c}
    \includegraphics[width=0.5\linewidth]{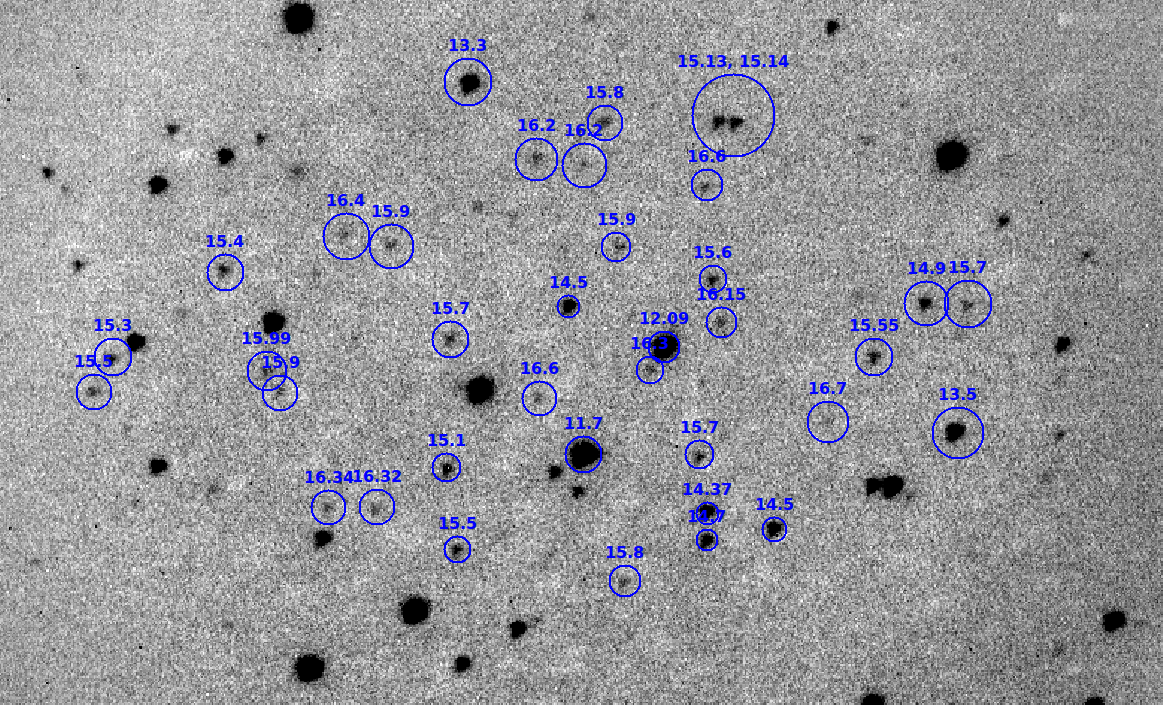}
\end{tabular}
\end{center}
    \caption
{\label{fig:12}Stars in the photometric standard field RU 149 marked with their respective magnitudes in J band.  }
\end{figure}

Landolt, A. U. in 1992 \cite {landolt92} listed photometric standards around the celestial equator whose 
V band magnitudes range between $\sim$ 11 - 16. RU 149 a blue star, and other surrounding stars (RU 149 A - G) 
in the field which are in RA 07 - 08 h was observed using NICSPol. Figure \ref {fig:12} shows the field observed and 
by comparing with the standard 2mass image, the magnitudes of the stars in J band were 
obtained which are marked in the figure. To estimate the 
limiting magnitude which could be achieved using NICSPol, polarization analysis of all the stars in the field was carried out.  
The exposure time per frame was 10 s (5 positions, 3 frames each, 4 angles) which resulted in a 
net exposure of 10 min for the complete observation.
It could be seen that with this net exposure it is possible to measure polarization for stars as faint as $\sim$15 mag in 
J.
Hence with a higher exposure it is possible to reach up to $\sim$16.5 (fainter stars seen in the field) using NICSPol.
This matches very well with the number quoted earlier in Section \ref {sec:3} ($\sim$16.8).
Since the observed field is closer to the Galactic plane the stars in the field can not be concluded as unpolarized. 

\section{Summary}
\label{sec:7}

We have presented here the design and test results of NICSPol, an NIR polarimeter add-on 
for NICS instrument covering wavelength range of 0.8 - 2.5 $\mu$m for the 1.2 m telescope at MIRO. 
NICSPol is the first imaging polarimeter in India and covers a FOV of 3.9 arc min dia 
and has a pixel scale of 0.5 arc sec per pixel. 
The polarization analysis is performed using IRAF and IDL following the standard 
Stokes methodology to obtain the degree of polarization and the polarization angle. Observations of 
NIR polarimetric standards as well as unpolarized stars using NICSPol show that the polarization 
degree and angle can be accurately determined within 1\% and few degrees, respectively.
This shows that NICSPol is suitable to carry out NIR polarization of sources with magnitudes
brighter than $\sim$16 over J, H, Ks bands and polarization greater than $\sim$1\%. 
As discussed in section \ref {sec:6.4}, for highly polarized sources, observations with large 
step angles would be sufficient, however for sources with low polarization, 
we may use more steps to fill the modulation curve and achieve higher accuracy.
NICSPol is a general purpose instrument at MIRO and can be used for observations of both point and 
extended sources. In future, we plan to use NICSPol simultaneously with the optical polarimeter 
available at the 50 cm telescope at Mt Abu to study wide variety of galactic and extra galactic sources
and hence many more scientific results are expected in near future.

\appendix    

\acknowledgments 
Contribution of PRL workshop is acknowledged for their support in making required parts and 
the overall support involved in mounting the WGP between the telescope and the
NICS instrument. 
We thank the observatory staff at MIRO for their technical contributions and for the support 
of the observations. We thank Mohan Lal for
his technical help during the prototype machining of WGP support for initial tests of NICSPol concept.
We also thank Aravind K and Prashanth Kumar Kasarla for their help in making the delrin mount for the polarizer. 
The authors thank the telescope operator trainees Prashant Chauhan and CM Mukesh
for their timely help during the observations.

\bibliographystyle{spiejour}   

\begin{table}[!ht]
\caption{Specifications of NICSPol}
\label{tab:1}       
\begin{center}
\begin{tabular}{|l|l|}
\hline
\rule[-1ex]{0pt}{3.5ex}  Parameter & Specifications \\
\hline\hline
\rule[-1ex]{0pt}{3.5ex}  Dimensions of WGP & WP25L-UB - 25.0 x 25.0 mm \\
\hline
\rule[-1ex]{0pt}{3.5ex}  Wavelength range & 250 nm to 4 $\mu$m (useful range for NICS : 0.8 to 2.5 $\mu$m\\
\hline
\rule[-1ex]{0pt}{3.5ex}  Available NIR bands & Y (0.97 - 1.07 $\mu$m), J (1.17 - 1.33 $\mu$m),\\
&H (1.49 - 1.78 $\mu$m), K (2.03 - 2.37 $\mu$m), Ks(1.99 - 2.31 $\mu$m)\\
\hline
\rule[-1ex]{0pt}{3.5ex}  Detector  & Teledyne H1RG detector with 1024 x 1024 arrays \\
\hline
\rule[-1ex]{0pt}{3.5ex}  Field of view & NICS - 8 x 8 sq arc min (imaging mode)\\
 & NICSPol - 3.9 arc min dia (polarimetric imaging mode)\\
\hline
\rule[-1ex]{0pt}{3.5ex}  Pixel scale &  0.5 arc sec per pixel\\
\hline
\rule[-1ex]{0pt}{3.5ex}  NICS optics and detector & Maintained at $\sim$ 77 K(Cooled by LN2)\\
\hline
\rule[-1ex]{0pt}{3.5ex}  WGP module & at ambient temperature\\
\hline
\rule[-1ex]{0pt}{3.5ex}  Limiting magnitudes   & $\sim$ 16.8 in JHKs \\
& Typically with 30 - 40 sec exposure per frame \\
& and a net exposure of 30 min \\
\hline
\end{tabular}
\end{center}
\end{table}

\begin{table}[!ht]
\caption{Log of all Observations using NICSPol}
\label{tab:2}       
\begin{center}
\begin{tabular}{|l|l|l|l|l|l|l|}
\hline
\rule[-1ex]{0pt}{3.5ex}  Source & Observed Nights & Filter & Positions & Frames & Orientations 
& Exposure/frame  \\
\hline\hline
\rule[-1ex]{0pt}{3.5ex}  HD 283809\textsuperscript{a} & 10 & J & 5 & 2 to 10 & 4 & 1 s \\
& 7 & H & 5 & 2 to 5 & 4 &  1 s\\
& 8 & Ks & 5 & 2 & 4 &  1 s\\
\hline
\rule[-1ex]{0pt}{3.5ex}  HD 204827\textsuperscript{a} & 5 & J & 5 & 3 & 4 &  801 ms\\
& 4 & H & 5 & 3 & 4 &  801 ms\\
\hline
\rule[-1ex]{0pt}{3.5ex}  HD 29333\textsuperscript{a} & 5 & J & 4 & 2 & 4 &  1 s \\
& 2 & H & 4 & 2 & 4 &  1 s\\
& 1 & Ks & 4 & 2 & 4 &  1 s\\
\hline
\rule[-1ex]{0pt}{3.5ex}  HD 283725\textsuperscript{a} & 2 & J & 5 & 3 & 4 &  1 s \\
\hline
\rule[-1ex]{0pt}{3.5ex}  HD 283855\textsuperscript{a} & 2 & J & 5 & 3 & 4 &  1 s \\
\hline
\rule[-1ex]{0pt}{3.5ex}  HD 283701\textsuperscript{a} & 1 & J & 5 & 3 & 4 &  1 s \\
\hline
\rule[-1ex]{0pt}{3.5ex}  HD 56591\textsuperscript{b} & 1 & J & 3 & 3 & 11 &  2 s \\
\hline
\rule[-1ex]{0pt}{3.5ex}  HD 26212\textsuperscript{b} & 1 & J & 3 & 3 & 4 &  2 s \\
& 1 & H & 3 & 3 & 4 &  2 s\\
& 1 & Ks & 3 & 3 & 4 &  2 s\\
& 1 & K & 3 & 3 & 4 &  2 s\\
& 1 & Y & 3 & 3 & 4 &  2 s\\
\hline
\rule[-1ex]{0pt}{3.5ex}  NGC 2548\textsuperscript{b} & 1 & J & 5 & 5 & 4 &  20 s \\
\hline
\rule[-1ex]{0pt}{3.5ex}  M3\textsuperscript{b} & 1 & J & 3 & 5 & 4 &  30 s \\
\hline
\rule[-1ex]{0pt}{3.5ex}  HD 202573\textsuperscript{c} & 2 & J & 5 & 2 & 4 &  1 s \\
& 2 & H & 5 & 2 & 4 &  1 s\\
& 2 & Ks & 5 & 2 & 4 &  1 s\\
& 2 & K & 5 & 3 & 4 &  1 s\\
\hline
\rule[-1ex]{0pt}{3.5ex}  HD 212311\textsuperscript{c} & 7 & J & 5 & 2 & 4 &  3 s \\
& 5 & H & 5 & 2 & 4 &  3 s\\
& 3 & Ks & 5 & 2 & 4 &  3 s\\
\hline
\rule[-1ex]{0pt}{3.5ex}  HD 103095\textsuperscript{c} & 3 & J & 4 to 8 & 5 & 4 &  801 ms \\
\hline
\rule[-1ex]{0pt}{3.5ex}  HD 65583\textsuperscript{c} & 1 & J & 7 & 5 & 4 &  801 ms \\
\hline
\rule[-1ex]{0pt}{3.5ex}  RA 07 - 08 h\textsuperscript{d} & 1 & J & 5 & 3 & 4 &  10 s \\
\hline
\end{tabular}
\end{center}
a - Polarized standards, b - Sources with WGP to check for 100 \% polarization, c - Unpolarized standards, d - Photometric standard star field.
\end{table}

\begin{table}[!ht]
\caption{Observed PF, PA for 100 \% polarized standard stars.}
\begin {center}
\label{tab:3}       
\begin{tabular}{|l|l|l|l|}
\hline
\rule[-1ex]{0pt}{3.5ex} Source & Band & Observed PF (\%) & Obs PA (\textsuperscript{o})   \\
\hline\hline
\rule[-1ex]{0pt}{3.5ex} HD 56591 & J &  100.75$\pm$0.52 &  -38.36$\pm$0.09  \\
\hline
\rule[-1ex]{0pt}{3.5ex} HD 26212 & J &  100.15$\pm$0.94 &  -38.45$\pm$0.24  \\
& H   & 100.16$\pm$0.66 &   -38.37$\pm$0.18  \\
& Ks   & 100.5$\pm$0.95 &   -38.21$\pm$0.24  \\
& K   & 100.74$\pm$0.96 &   -38.20$\pm$0.20  \\
& Y   & 100.58$\pm$1.22 &   -38.23$\pm$0.3  \\
\hline
\end{tabular}
\end{center}
\end{table}

\begin{table}[!ht]
\caption{Observed PF and PA for 100 \% polarized light for the source NGC 2548.}
\begin {center}
\label{tab:4}       
\begin{tabular}{|l|l|l|}
\hline
\rule[-1ex]{0pt}{3.5ex} Source & Obs PF (\%) & Obs PA (\textsuperscript{o}) \\
\hline \hline
\rule[-1ex]{0pt}{3.5ex} Star 1 & 100.45$\pm$0.62 & 15.24$\pm$0.16\\
\hline
\rule[-1ex]{0pt}{3.5ex} Star 2 & 99.96$\pm$0.8 & 15.29$\pm$0.18\\
\hline
\rule[-1ex]{0pt}{3.5ex} Star 3 & 100.66$\pm$1.67 & 15.43$\pm$0.39\\
\hline
\rule[-1ex]{0pt}{3.5ex} Star 4 & 100.22$\pm$0.87 & 15.4$\pm$0.22\\
\hline
\rule[-1ex]{0pt}{3.5ex} Star 5 & 99.66$\pm$0.36 & 15.49$\pm$0.10\\
\hline
\rule[-1ex]{0pt}{3.5ex} Star 6 & 100.53$\pm$1.4 & 15.97$\pm$0.33\\
\hline
\rule[-1ex]{0pt}{3.5ex} Star 7 & 100.44$\pm$0.8 &  15.53$\pm$0.21\\
\hline
\rule[-1ex]{0pt}{3.5ex} Star 8 & 101.16$\pm$2.76 & 15.62$\pm$0.63\\
\hline
\rule[-1ex]{0pt}{3.5ex} Star 9 & 99.57$\pm$0.27 & 15.44$\pm$0.07\\
\hline
\rule[-1ex]{0pt}{3.5ex} Star 10 & 96.87$\pm$0.61 & 15.41$\pm$0.18\\
\hline
\rule[-1ex]{0pt}{3.5ex} Star 11 & 96.12$\pm$1.37 & 14.21$\pm$0.35\\
\hline
\rule[-1ex]{0pt}{3.5ex} Star 12 & 101.57$\pm$2.48 & 15.24$\pm$0.57\\
\hline
\rule[-1ex]{0pt}{3.5ex} Star 13 & 100.9$\pm$1.15 & 15.49$\pm$0.29\\
\hline
\rule[-1ex]{0pt}{3.5ex} Star 14 & 100.82$\pm$1.41 & 15.59$\pm$0.34\\
\hline
\rule[-1ex]{0pt}{3.5ex} Star 15 & 99.04$\pm$3.63 & 14.88$\pm$0.83\\
\hline
\rule[-1ex]{0pt}{3.5ex} Star 16 & 96.02$\pm$2.95 & 20.69$\pm$0.76\\
\hline
\rule[-1ex]{0pt}{3.5ex} Star 17 & 98.00$\pm$2.85 & 18.57$\pm$0.73\\
\hline
\rule[-1ex]{0pt}{3.5ex} Star 18 & 100.92$\pm$2.48 & 15.04$\pm$0.57\\
\hline
\end{tabular}
\end{center}
\end{table}

\begin{table}[!ht]
\caption{Observed PF for Unpolarized standard stars.}
\begin {center}
\label{tab:5}       
\begin{tabular}{|l|l|l|l|}
\hline
\rule[-1ex]{0pt}{3.5ex} Source & Band & Observed PF (\%)  & Ref  \\
\hline\hline
\rule[-1ex]{0pt}{3.5ex}  HD 202573 & J &  1.2$\pm$0.24  & \cite {serkowski74}\\
& K  & 0.21$\pm$0.23 &\\
\hline
\rule[-1ex]{0pt}{3.5ex}  HD 212311 & J & 0.65$\pm$0.68  & \cite {turnshek90}\\
& H  & 0.52$\pm$0.49 &\\
& Ks  & 1.37$\pm$0.81 &\\
\hline
\rule[-1ex]{0pt}{3.5ex} HD 103095 & J  & 0.39$\pm$0.23   & \cite {serkowski74}\\
\hline
\rule[-1ex]{0pt}{3.5ex} HD 65583 & J  & 0.07$\pm$0.52   & \cite {serkowski74}\\
\hline
\end{tabular}
\end{center}
\end{table}

\begin{table}[!ht]
\caption{Standard PF, PA and Observed PF, PA for Polarized standard stars.}
\begin {center}
\label{tab:6}       
\begin{tabular}{|l|l|l|l|l|l|l|}
\hline
\rule[-1ex]{0pt}{3.5ex} Source & Band & Std PF (\%) & Obs PF (\%) & Std PA (\textsuperscript{o}) & Obs PA (\textsuperscript{o}) & Ref  \\
\hline\hline
\rule[-1ex]{0pt}{3.5ex}  HD 283809 & J &  3.81$\pm$0.07 & 3.19$\pm$0.69 & 57$\pm$1 & 49.52$\pm$6.2 & \cite {whittet92}\\
& H &  2.59$\pm$0.07 & 2.12$\pm$0.33 & 58$\pm$1 & 56.29$\pm$4.4 & \cite {whittet92}\\
& Ks &  1.71$\pm$0.11 & 1.53$\pm$0.49 & 55$\pm$1 & 59.04$\pm$9.18 & \cite {whittet92}\\
\hline
\rule[-1ex]{0pt}{3.5ex} HD 204827 & J & 2.83$\pm$0.07 & 2.12$\pm$0.49 & 61.1$\pm$0.8 & 63.85$\pm$6.55 & \cite {schulz83}\\
& H &   & 1.71 $\pm$0.33 &  & 60.45$\pm$5.47&\\
\hline
\rule[-1ex]{0pt}{3.5ex} HD 29333 & J & 2.88$\pm$0.03  & 2.47$\pm$0.49 & 69$\pm$1 & 69.2$\pm$5.67 & \cite {whittet92}\\
& H & 1.81$\pm$0.04  & 1.81$\pm$0.49 & 68$\pm$1 & 70.57$\pm$7.73 & \cite {whittet92}\\
& Ks & 1.19$\pm$0.08  & 1.14$\pm$0.49 & 73$\pm$1 & 66.46$\pm$12.3 & \cite {whittet92}\\
\hline
\rule[-1ex]{0pt}{3.5ex} HD 283725 & J & 2.5$\pm$0.3  & 2.49$\pm$0.33 & 66$\pm$1 & 62.44$\pm$5.61 & \cite {whittet92}\\
\hline
\rule[-1ex]{0pt}{3.5ex} HD 283855 & J & 2.58$\pm$0.03  & 2.27$\pm$0.33 & 46$\pm$1 & 172.16$\pm$14.4 & 
\cite {whittet92}\\
\hline
\rule[-1ex]{0pt}{3.5ex} HD 283701 & J & 1.68$\pm$0.12  & 1.65$\pm$0.33 & 33$\pm$1.3 & 7.16$\pm$11.31 & \cite {whittet01}\\
\hline
\end{tabular}
\end{center}
\end{table} 


\begin{thebibliography}{10}

\bibitem{aitken88}
D.~K. {Aitken}, ``{Infrared polarimetry},'' {\em Vistas in Astronomy} {\bf 31},
  755--762  (1988).

\bibitem{gnedin78}
Y.~N. {Gnedin}, G.~G. {Pavlov}, and Y.~A. {Shibanov}, ``{The effect of vacuum
  birefringence in a magnetic field on the polarization and beaming of X-ray
  pulsars},'' {\em Soviet Astronomy Letters} {\bf 4}, 117--119  (1978).

\bibitem{meszaros80}
P.~{Meszaros}, W.~{Nagel}, and J.~{Ventura}, ``{Exact and approximate solutions
  for the one-dimensional transfer of polarized radiation, and applications to
  X-ray pulsars},'' {\em Astrophysical Journal} {\bf 238}, 1066--1080  (1980).

\bibitem{martin75}
P.~G. {Martin}, ``{Some implications of 10-micron interstellar polarization},''
  {\em Astrophysical Journal} {\bf 202}, 393--399  (1975).

\bibitem{kandori06}
R.~{Kandori}, N.~{Kusakabe}, M.~{Tamura}, {\em et~al.}, ``{SIRPOL: a JHK
  $_{s}$-simultaneous imaging polarimeter for the IRSF 1.4-m telescope},'' in
  {\em Society of Photo-Optical Instrumentation Engineers (SPIE) Conference
  Series},  {\em Society of Photo-Optical Instrumentation Engineers (SPIE)
  Conference Series} {\bf 6269}, 626951  (2006).

\bibitem{clemens07}
D.~P. {Clemens}, D.~{Sarcia}, A.~{Grabau}, {\em et~al.}, ``{Mimir: A
  Near-Infrared Wide-Field Imager, Spectrometer and Polarimeter},'' {\em
  Publications of the astronomical society of the pacific} {\bf 119},
  1385--1402  (2007).

\bibitem{schultz03}
A.~B. {Schultz}, M.~{Sosey}, L.~M. {Mazzuca}, {\em et~al.}, ``{Post-NCS
  Performance of the HST NICMOS},'' in {\em IR Space Telescopes and
  Instruments},  J.~C. {Mather}, Ed., {\em Society of Photo-Optical
  Instrumentation Engineers (SPIE) Conference Series} {\bf 4850}, 858--866
  (2003).

\bibitem{jones07}
T.~J. {Jones} and C.~C. {Packham}, ``{MMTPol, A High Precision Imaging
  Polarimeter for the MMT},'' in {\em American Astronomical Society Meeting
  Abstracts},  {\em Bulletin of the American Astronomical Society} {\bf 39},
  748  (2007).

\bibitem{wolf02}
S.~{Wolf}, L.~{Vanzi}, and N.~{Ageorges}, ``{Technical report on the
  polarimetric mode of SOFI}.''
  \url{https://www.eso.org/sci/facilities/lasilla/instruments/sofi/doc/manual/sofiman_1.6.pdf}
   (2002).
\newblock [Online; accessed 18-Oct-2018].

\bibitem{devaraj18}
R.~{Devaraj}, A.~{Luna}, L.~{Carrasco}, {\em et~al.}, ``{POLICAN: A
  Near-infrared Imaging Polarimeter at the 2.1m OAGH Telescope},'' {\em
  Publications of the astronomical society of the pacific} {\bf 130}, 055002
  (2018).

\bibitem{watanbe05}
M.~{Watanabe}, H.~{Nakaya}, T.~{Yamamuro}, {\em et~al.}, ``{TRISPEC: A
  Simultaneous Optical and Near-Infrared Imager, Spectrograph, and
  Polarimeter},'' {\em Publications of the astronomical society of the pacific}
  {\bf 117}, 870--884  (2005).

\bibitem{beuzit13}
J.-L. {Beuzit}, ``{SPHERE: a Planet Finder Instrument for the VLT},'' {\em
  European Planetary Science Congress} {\bf 8}, EPSC2013--954  (2013).

\bibitem{akitaya14}
H.~{Akitaya}, Y.~{Moritani}, T.~{Ui}, {\em et~al.}, ``{HONIR: an optical and
  near-infrared simultaneous imager, spectrograph, and polarimeter for the
  1.5-m Kanata telescope},'' in {\em Ground-based and Airborne Instrumentation
  for Astronomy V},  {\em Society of Photo-Optical Instrumentation Engineers
  (SPIE) Conference Series} {\bf 9147}, 91474O  (2014).

\bibitem{anandrao08}
B.~{Anandarao}, E.~H. {Richardson}, A.~{Chakraborty}, {\em et~al.}, ``{A
  wide-field near-infrared camera and spectrograph for the Mt. Abu 1.2 m
  telescope},'' in {\em Ground-based and Airborne Instrumentation for Astronomy
  II},  {\em Society of Photo-Optical Instrumentation Engineers (SPIE)
  Conference Series} {\bf 7014}, 70142Y  (2008).

\bibitem{kolokolova15}
L.~{Kolokolova}, J.~{Hough}, and A.-C. {Levasseur-Regourd}, {\em {Polarimetry
  of Stars and Planetary Systems}}  (2015).

\bibitem{trippe14}
S.~{Trippe}, ``{[Review] Polarization and Polarimetry},'' {\em Journal of
  Korean Astronomical Society} {\bf 47}, 15--39  (2014).

\bibitem{snik13}
F.~{Snik} and C.~U. {Keller}, {\em {Astronomical Polarimetry: Polarized Views
  of Stars and Planets}}, 175  (2013).

\bibitem{hines09}
D.~C. {Hines}, C.~C. {Packham}, A.~{Adamson}, {\em et~al.}, ``{O/IR Polarimetry
  for the 2010 Decade (CGT): Science at the Edge, Sharp Tools for All},'' in
  {\em astro2010: The Astronomy and Astrophysics Decadal Survey},  {\em ArXiv
  Astrophysics e-prints} {\bf 2010}  (2009).

\bibitem{hoffman09}
J.~L. {Hoffman}, D.~C. {Hines}, A.~{Adamson}, {\em et~al.}, ``{O/IR Polarimetry
  for the 2010 Decade (SSE): Science at the Edge, Sharp Tools for All},'' in
  {\em astro2010: The Astronomy and Astrophysics Decadal Survey},  {\em ArXiv
  Astrophysics e-prints} {\bf 2010}  (2009).

\bibitem{clemens09}
D.~{Clemens}, B.-G. {Andersson}, A.~{Adamson}, {\em et~al.}, ``{O/IR
  Polarimetry for the 2010 Decade (PSF): Science at the Edge, Sharp Tools for
  All},'' in {\em astro2010: The Astronomy and Astrophysics Decadal Survey},
  {\em ArXiv Astrophysics e-prints} {\bf 2010}  (2009).

\bibitem{clemens09gan}
D.~{Clemens}, B.-G. {Andersson}, A.~{Adamson}, {\em et~al.}, ``{O/IR
  Polarimetry for the 2010 Decade (GAN): Science at the Edge, Sharp Tools for
  All},'' in {\em astro2010: The Astronomy and Astrophysics Decadal Survey},
  {\em ArXiv Astrophysics e-prints} {\bf 2010}  (2009).

\bibitem{clemens12}
D.~P. {Clemens}, A.~F. {Pinnick}, M.~D. {Pavel}, {\em et~al.}, ``{The Galactic
  Plane Infrared Polarization Survey (GPIPS)},'' {\em Astrophysical Journal}
  {\bf 200}, 19  (2012).

\bibitem{ganeshs13}
S.~{Ganesh}, K.~S. {Baliyan}, S.~{Chandra}, {\em et~al.}, ``{Automated
  telescope for variability studies},'' in {\em Astronomical Society of India
  Conference Series},  {\em Astronomical Society of India Conference Series}
  {\bf 9}  (2013).

\bibitem{vadawale15}
S.~V. {Vadawale}, T.~{Chattopadhyay}, A.~R. {Rao}, {\em et~al.}, ``{Hard X-ray
  polarimetry with Astrosat-CZTI},'' {\em Astronomy \& Astrophysics} {\bf 578}
  (2015).

\bibitem{wpp}
``{Wollastron Polarizing Prisms}.''
  \url{https://www.newport.com/f/wollaston-polarizing-prisms}.

\bibitem{wdw}
E.~{Oliva}, ``{Wedged double Wollaston, a device for single shot polarimetric
  measurements}.''
  \url{http://citeseerx.ist.psu.edu/viewdoc/download?doi=10.1.1.868.880&rep=rep1&type=pdf}
   (1996).

\bibitem{wgp}
``{Wire Grid Polarizers}.''
  \url{https://www.nilt.com/files/pdf/app_note_wire_grid_polarizers.pdf}
  (2009).

\bibitem{thorlabs}
``{Thorlabs}.''
  \url{https://www.thorlabs.com/newgrouppage9.cfm?objectgroup_id=5510}.

\bibitem{stokes52}
G.~G. {Stokes}, ``{On the Composition and Resolution of Streams of Polarized
  Light from different Sources},'' {\em Transactions of the Cambridge
  Philosophical Society} {\bf 9}, 399  (1851).

\bibitem{chandra60}
S.~{Chandrasekhar}, {\em {Radiative transfer}}  (1960).


\bibitem{frecker76}
J.~E. {Frecker} and K.~{Serkowski}, ``{Linear polarimeter with rapid
  modulation, achromatic in the 0.3-1.1-micron range},'' {\em Applied Optics}
  {\bf 15}, 605  (1976).

\bibitem{whittet92}
D.~C.~B. {Whittet}, P.~G. {Martin}, J.~H. {Hough}, {\em et~al.}, ``{Systematic
  variations in the wavelength dependence of interstellar linear
  polarization},'' {\em Astrophysical Journal} {\bf 386}, 562--577  (1992).

\bibitem{serkowski74}
K.~{Serkowski}, ``{Polarimeters for Optical Astronomy},'' in {\em IAU Colloq.
  23: Planets, Stars, and Nebulae: Studied with Photopolarimetry},
  T.~{Gehrels}, Ed., 135  (1974).

\bibitem{whittet01}
D.~C.~B. {Whittet}, P.~A. {Gerakines}, J.~H. {Hough}, {\em et~al.},
  ``{Interstellar Extinction and Polarization in the Taurus Dark Clouds: The
  Optical Properties of Dust near the Diffuse/Dense Cloud Interface},'' {\em
  Astrophysical Journal} {\bf 547}, 872--884  (2001).

\bibitem{landolt92}
A.~U. {Landolt}, ``{UBVRI photometric standard stars in the magnitude range
  11.5-16.0 around the celestial equator},'' {\em Astronomical Journal} {\bf
  104}, 340--371  (1992).

\bibitem{turnshek90}
D.~A. {Turnshek}, R.~C. {Bohlin}, R.~L. {Williamson}, II, {\em et~al.}, ``{An
  atlas of Hubble Space Telescope photometric, spectrophotometric, and
  polarimetric calibration objects},'' {\em Astrophysical Journal} {\bf 99},
  1243--1261  (1990).

\bibitem{schulz83}
A.~{Schulz} and R.~{Lenzen}, ``{New polarization measurements of HD 183143, HD
  204827, and CYG OB 2 Sch. No. 12},'' {\em AAP} {\bf 121}, 158--161  (1983).

\end{thebibliography}
\end{document}